# PolyDebug: A Framework for Polyglot Debugging

Philémon Houdaille[a,b,c] , Djamel Eddine Khelladi[b,a] , Benoit Combemale[a] , Gunter Mussbacher[d,c] , and Tijs van der Storm[e,f]

a  University of Rennes, Rennes, France
b  CNRS, France
c  Inria, France
d  McGill University, Montréal, Canada
e  Centrum Wiskunde & Informatica (CWI), Amsterdam, The Netherlands
f  University of Groningen, Groningen, The Netherlands

**Abstract**     As software grows increasingly complex, the quantity and diversity of concerns to be addressed also rises. To answer this diversity of concerns, developers may end up using multiple programming languages in a single software project, a practice known as polyglot programming. This practice has gained momentum with the rise of execution platforms capable of supporting polyglot systems.

However, despite this momentum, there is a notable lack of development tooling support for developers working on polyglot programs, such as in debugging facilities. Not all polyglot execution platforms provide debugging capabilities, and for those that do, implementing support for new languages can be costly.

This paper addresses this gap by introducing a novel debugger framework that is language-agnostic yet leverages existing language-specific debuggers. The proposed framework is dynamically extensible to accommodate the evolving combination of languages used in polyglot software development. It utilizes the Debug Adapter Protocol (DAP) to integrate and coordinate existing debuggers within a debugging session.

We found that using our approach, we were able to implement polyglot debugging support for three different languages with little development effort. We also found that our debugger did not introduce an overhead significant enough to hinder debugging tasks in many scenarios; however performance did deteriorate with the amount of polyglot calls, making the approach not suitable for every polyglot program structure.

The effectiveness of this approach is demonstrated through the development of a prototype, *PolyDebug*, and its application to use cases involving C, JavaScript, and Python. We evaluated *PolyDebug* on a dataset of traditional benchmark programs, modified to fit our criteria of polyglot programs. We also assessed the development effort by measuring the source lines of code (SLOC) for the prototype as a whole as well as its components.

Debugging is a fundamental part of developing and maintaining software. Lack of debug tools can lead to difficulty in locating software bugs and slow down the development process. We believe this work is relevant to help provide developers proper debugging support regardless of the runtime environment.

### The Art, Science, and Engineering of Programming

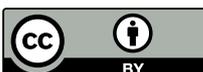

**PolyDebug: A Framework for Polyglot Debugging**

## 1 Introduction

As complexity of software increases, developers have to face more and more challenges when developing applications. One of these challenges is the diversity of concerns encountered: to achieve its end goal, a single program might present a variety of different concerns, such as machine learning algorithms, network communications, video encoding, user interface, *etc*. A plethora of programming languages exist today to support the increasingly complex software engineering needs across diverse domains. The activity of writing code using multiple programming languages in conjunction has been defined as polyglot programming [3, 4, 7, 10, 20, 32].

Indeed, different concerns are not always best answered in the same manner, and one factor reducing the accidental complexity is the programming language being used. For example, Python is well suited for data management due to its library ecosystem, while high performance video encoding is better realized in a language with lower-level control of memory such as C. Polyglot programming exploits this concept with the idea that for the best possible language to be used in answer to a specific concern, the part of the program written in this language needs to be able to freely interact with parts written in other languages that will solve different problems in the same program.

Examples of execution methods supporting polyglot programming include MetaCall [19], Foreign Function Interfaces [17] (FFI), GraalVM [32] and its Truffle [31] language framework, or WebAssembly [9]. There is a high diversity of such run-time environments, with varying syntax, semantics, and implementations. Regardless of the underlying technological stack and different implementation choices, we refer to any tool supporting the execution of polyglot programs as a *polyglot execution platform*, and use this term in the rest of the paper.

Despite the momentum and high promise of polyglot programming, there is a notable lack of development tooling support for developers working on polyglot programs, such as for debugging activities. Indeed developing run-time support for polyglot programs is already a challenge of its own, and adding debug support on top of that often falls out of scope for polyglot platforms. Nonetheless, some solutions exist to reduce the engineering effort of providing such debugging capabilities such as the DWARF[1] format for intermediate representation-based platforms (*e.g.* WebAssembly), or the GraalVM debugger [28] in the context of the Graal and Truffle ecosystems. However these solutions still require a fair amount of implementation effort, respectively in implementing a debugger for the intermediate representation or in implementing a language with the dedicated Truffle library.

This paper proposes a novel framework to provide polyglot debugging support in a generic manner, agnostic of both languages and polyglot execution platforms. Our approach leverages existing language-specific debuggers and coordinates these debuggers. This allows us to implement behaviour that attempts to emulate the behaviour of an existing polyglot execution platform, thus enabling debugging of programs written for that platform. The proposed architecture is dynamically extensible to accommodate the evolving combination of languages used in polyglot software development. We achieve these features partly by relying on the Debug Adapter Protocol (DAP). Today, many debuggers are built by implementing DAP.[2] The main advantage offered by DAP is that the debugger (*i.e.* the DAP server) becomes agnostic of the client. In particular, it promotes the reuse of debuggers between IDEs by providing a standardized interface to debugging actions. Thus, our approach enables the integration and coordination of existing debuggers of *n* different programming languages. Developers can decide to mix any programming language, and our approach can provide debugging support by leveraging DAP-compliant debuggers. Polyglot platforms can provide both execution flow and data flow constructs [11]. While our approach can in theory support both of these, this paper focuses mostly on supporting execution flow during a debug session. We elaborate more on limitations and potential solutions related to data flow in Section 4 but mostly leave such solutions to future work.

---

[1] https://dwarfstd.org/ (Visited on 2025-01-21).

The effectiveness of our approach is demonstrated through the development of a prototype emulating simplified GraalVM polyglot semantics, *PolyDebug*, and its application to use cases of polyglot programs involving C, JavaScript, and Python. We evaluated *PolyDebug* on a dataset of three algorithms taken from the Benchmarks Game [8], a benchmark comparing the performance of several programming languages. In fact, the benchmark consists of the implementation of the same programs in different languages. Each program in each language is implemented through different functions, *e.g.* f1(), f2(), f3(), etc. We made those programs polyglot by simply switching the implementation language of one or multiple functions. For example, we could call f1() in its original source language C and the f2() from the Python implementation and create a polyglot program that is semantically equivalent to the original.

Results show that with just over a thousand lines of code of implementation, *PolyDebug* is able to execute and debug polyglot programs spanning three programming languages. This represents a relatively low implementation effort, although there is an additional engineering effort related to proficiency with the target language that is difficult to measure quantitatively. *PolyDebug* performance varies depending on the structure of the program being debugged, more specifically in where polyglot calls are made; in the case of infrequent calls, overhead is barely noticeable, but this overhead can explode in some corner cases where a program has many polyglot calls to small program parts (*i.e.* the trade-off between the overhead of performing a polyglot call and the execution time of the target of the call becomes not worth it).

To summarize, our novel contributions are as follows:

- A new framework to provide support for polyglot debugging.
- An instance of this framework through the *PolyDebug* prototype, supporting C, JavaScript, and Python with a simplified model of GraalVM polyglot semantics.

---

[2] https://microsoft.github.io/debug-adapter-protocol/implementors/adapters/ (Visited on 2025-01-21).

- An evaluation on a dataset of polyglot programs based on three algorithms from the Benchmarks Game [8].

The rest of the paper is structured as follows. We cover background topics required to understand our approach in Section 2. Then, we present our proposed architecture in Section 3 and the execution model of the framework in Section 4, and detail our prototype implementation of this architecture in Section 5. We evaluate both our approach and the corresponding implementation in Section 6, present relevant work related to our own in the literature in Section 7, and finally conclude the paper in Section 8 while also providing some perspectives on directions for future work.

## 2 Background

This section covers the background topics required to discuss our approach. It first covers the subject of polyglot programming and the state of its associated tooling, and then discusses the Debug Adapter Protocol.

### 2.1 Polyglot Programming and Tooling

Polyglot programming refers to the activity of developing a program made up of different programming languages [20]. To support this practice, a few technologies have been developed in order to provide a runtime environment for polyglot programs. We refer to these technologies as polyglot execution platforms. Examples of polyglot platforms include the Java Virtual Machine, .NET,[3] GraalVM [32], WebAssembly [9], or MetaCall.[4] These platforms mostly use a unified runtime, with an intermediate representation that languages are compiled to before being actually executed. However, this is not a restriction of our scope: it is also possible to implement polyglot platforms through a more component-based approach, with interfaces serving as the interaction point between parts of the program written in different languages. Regardless, this still qualifies as a polyglot platform in our definition, so long as the programmer needs to interact with multiple languages during a development session. While there are many industrial and academic efforts investigating polyglot platforms, this is less so the case for development tooling [18].

One solution that could help to provide polyglot debugging tools is the DWARF[5] format, which is an extensible format to link source code and compiled code in debugging tasks. However, this is only applicable to polyglot platforms that use a single compiled representation for all languages.

Another existing solution is Visual Studio's mixed mode debugging,[6] which is a built-in debugger mode allowing the navigation of native and managed code in a single session. However, this is a solution that lacks genericity as it seems only available in the Visual Studio IDE, and not easily extensible to languages outside the .NET ecosystem.

---

[3] https://dotnet.microsoft.com/ (Visited on 2025-01-21).
[4] https://metacall.io/ (Visited on 2025-01-21).
[5] https://dwarfstd.org/ (Visited on 2025-01-21).
[6] https://learn.microsoft.com/en-us/visualstudio/debugger/how-to-debug-in-mixed-mode (Visited on 2025-01-21).

Many research efforts [12, 13, 22, 26] on supporting the activity of polyglot programming are based on the GraalVM ecosystem, which provides a powerful API for runtime instrumentation [28]. This is realized by leveraging Truffle, a library to implement interpreters. By implementing a language with Truffle, this language is then able to run on GraalVM and benefit from many of its features, such as the ability to interact with other Truffle languages. This is the mechanism that allows for polyglot programming.

Figure 1 showcases an example of a small two-part GraalVM program to illustrate how polyglot programs are written. On the left side is the Python part of the program which is also the entry point, and on the right side is the Javascript part. Line 2 of foo.py calls the GraalVM polyglot.export function, which makes the value 42 available under the name x for all subsequent parts of the program. Line 3 then calls the execution of foo.js with the polyglot.eval function. foo.js starts by doing the reverse operation to export at line 2, with the Polyglot.import call on x. This stores the value 42 in the local variable y, which is then logged to the Javascript console (*i.e.* printed). The Javascript program then ends at line 4 by stating the value 7. This becomes the return value of the polyglot.eval call at line 3 of foo.py, and thus the value 7 is stored in the local variable foo, which is then printed. This example is simple for illustration and brevity purposes, but showcases the three main functions of GraalVM's polyglot API (*i.e.* eval, export, and import).

In the rest of this paper, we instantiate our approach to a simplified version of the GraalVM polyglot API, which gives a reference of real polyglot syntax and semantics to illustrate feasibility. We thus introduce the polyglotEval function, modeling a simplified behaviour of Graal's `polyglot.eval` where return values are passed by value rather than by reference. It is worth noting here that our approach can only target polyglot platforms where the construct used to invoke polyglot interactions are syntactically valid in the base language (*i.e.* `polyglot.eval` raises no issues, but an hypothetical construct such as "@#eval" would use tokens potentially invalid in the grammar of many languages).

The Truffle library also provides an instrumentation API that allows defining tools generically for any Truffle language. For instance, the GraalVM debugger is implemented in this manner, and is thus available to any language as soon as they are implemented using Truffle. However, implementing languages in this way is not without issues. Languages need to be implemented from scratch, a long process which can be error-prone, and can feel redundant for languages that already have their own non-polyglot implementation and ecosystem. Additionally, the GraalVM debugger is also constrained to the GraalVM part of the program; it cannot be used in conjunction with parts of the program using other technologies for polyglot interactions.

For this reason, we advocate for additional research on polyglot tooling techniques that abstract away from the execution platform, and can be applied regardless of the implementation of the runtime technology. We also advocate for polyglot programming techniques to rely on the decades of tooling developed for individual languages when

**PolyDebug: A Framework for Polyglot Debugging**

possible, as this tooling has often been highly optimized and does not always need to be re-implemented.

```
1  # foo.py
2  polyglot.export("x", 42)
3  foo = polyglot.eval("js", "foo.js")
4  print(foo) # prints 7
```

```
1  // foo.js
2  var y = Polyglot.import("x");
3  console.log(y); // prints 42
4  7;
```

**Figure 1** Example of a Python-Javascript GraalVM program that prints 42, then 7

## 2.2 Debug Adapter Protocol (DAP)

The Debug Adapter Protocol[7] (henceforth DAP) is a client-server protocol that aims to streamline the implementation of debug support in code editors. In DAP, the server is a debugger for a given language. Traditionally, the client is an integrated development environment (IDE), often implemented through features, such as modules or extensions. The idea behind DAP is that a given debug server only needs to be implemented once, and can then be reused to implement debug capabilities for its target language in many IDEs.

To this end, DAP provides a standardized interface. This interface allows the client and server to communicate through the concepts of *requests, responses,* and *events*. *Requests* are messages sent by the client to the server; *responses* are messages sent by the server to the client in reaction to a received request. One example of a request and response pair is the evaluate request, which asks the DAP server to evaluate the value of an expression. The response is the result of the evaluation. *Events* are messages sent by the server to the client that are not directly the result of responding to a request. One example of an event is the stopped event, which occurs when program execution is halted after stepping, hitting a breakpoint, or any other similar circumstance.

As DAP is the *de facto* standard method of implementing debug support for many IDEs, it benefits from a large set of readily available servers covering many different programming languages. Additionally, there has been research work on how to support a subset of DAP for DSLs [6]. All of this makes it an attractive option when trying to build a polyglot debugging solution, while reusing single-language tooling as much as possible.

As part of our approach, we assume a DAP server provides support for the language without any restriction on operations (specifically, not on operations related to reflexivity or other introspection mechanisms). We also assume that the DAP server supports the following DAP requests: launch, setBreakpoint, continue, stacktrace, next, and setExpression or setVariable.

---

[7] https://microsoft.github.io/debug-adapter-protocol/ (Visited on 2025-01-21).

## 3 Framework Architecture

This section describes the overall framework's architecture. We first give a broad overview, then detail the role of the components as well as how they can be specialized. We illustrate this specialization through our example of simplified GraalVM polyglot API.

### 3.1 Overview

The goal of our approach is to enable debugging of polyglot programs. We assume there exists an hypothetical client to which we want to expose a DAP server interface, so that they are able to perform standard debug operations according to DAP capabilities.

Our idea is to reuse language-specific DAP servers to defer any language-specific execution and debug operations directly to the relevant server. Polyglot execution can thus be realized by pausing debuggers at polyglot call locations, and switching to the next relevant debugger to continue executing the program.

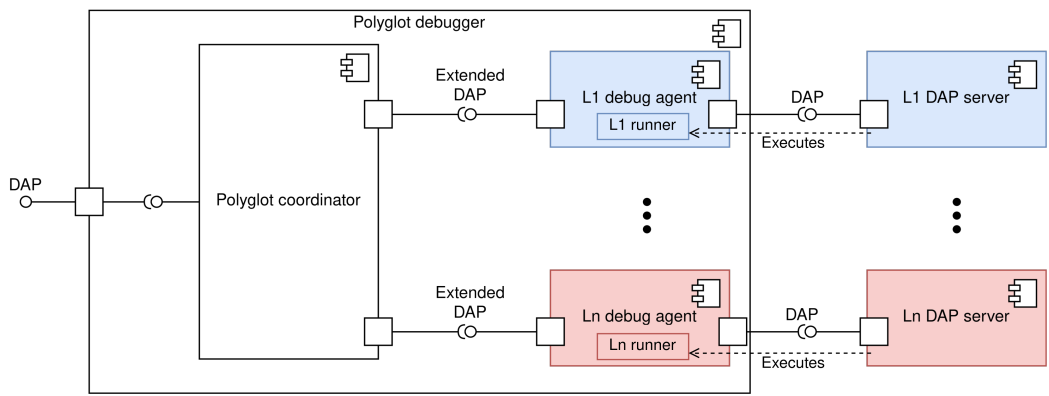

**Figure 2** *PolyDebug* overall architecture

The proposed extensible and language-agnostic architecture for polyglot debugging introduces a polyglot debugger (see center of Figure 2) as an intermediary between the client and the language-specific DAP servers. The polyglot debugger consists of two main components: (a) one polyglot coordinator, instantiating the execution model of the polyglot platform and (b) a set of debug agents each instantiating a supported language in the polyglot platform. The polyglot debugger can evolve over time by being extended with new debug agents.

The polyglot coordinator re-exposes DAP so that a client can interact with the polyglot debugger like with any other DAP server. Furthermore, the polyglot coordinator redirects the incoming DAP requests from the client to the appropriate debug agents, and is in charge of orchestrating the execution flow of the debugging session.

A debug agent interacts with an existing DAP server of a language, *i.e.* the proposed architecture allows the direct reuse of these servers without any modification of their implementation. The goal of the debug agent and its associated runner is to manage the DAP server so that program execution can be paused and resumed according to

**PolyDebug: A Framework for Polyglot Debugging**

requests from the polyglot coordinator. The agent re-exposes to the coordinator the classic DAP interface of the managed server, with a few additional operations.

Figure 3 shows the setup phase for a debug agent which is done once per agent when they are first launched. Agents can be launched during initialization of the polyglot debugger, as it requires a set of debug agents to perform debugging tasks. However, this set is not static: our approach also allows a plug-and-play strategy, where agents are dynamically added to the polyglot debugger as needed over the course of a debug session. The setup phase for adding an agent is identical regardless of when they are added to the set of agents available to the polyglot debugger. The agent first initializes its corresponding DAP server, with language-specific arguments. The DAP server is then tasked to launch the polyglot runner program, and to set appropriate breakpoints to handle polyglot operations. The DAP server then goes into a "standby" state, where it awaits further instructions from the debug agent to execute incoming code. We further detail this code execution mechanism and the role of the runner in Section 3.3.

While the polyglot coordinator is entirely agnostic of languages and can generically handle debug agents, the debug agents themselves target specific languages. Thus, they need to be implemented specifically for their target language. Nonetheless, they share a common role, namely: *1)* managing the code execution capabilities of the runner, *2)* controlling the flow of execution to allow polyglot interactions to occur through the polyglot coordinator, and *3)* exposing a DAP interface with the extension of the execute and setResult requests. We now detail the language-agnostic aspects required of the debug agents to perform these roles.

**3.2 Debug Agents**

The main role of a debug agent is to provide a bridge between a language specific DAP server and the generic polyglot coordinator. To this end, each debug agent re-exposes the classic DAP API to the coordinator, with additional requests as required for emulating the polyglot behaviour of the target platform. In the case of our polyglotEval function, these are the setResult and execute requests. The setResult request is simply a way to signal to a debug agent that it can resume execution, with the last polyglot call's return value being passed as argument. We detail its role in Section 3.3.

The execute request asks a debug agent to execute a provided part of a program as if continuing a standard debug session. This request is the main mechanism allowing the polyglot coordinator to appropriately redirect execution flow towards a corresponding debug agent.

In the regular DAP, a similar request is the evaluate request. However, we observed that its behaviour varied between different DAP servers, and most notably did not always take into account breakpoints when execution was involved in the evaluation. Thus, we could not use it as it is, and created the execute request.

As mentioned earlier, another role of debug agents is to handle starting, pausing and resuming execution for each DAP server. However, we cannot simply start, pause and restart the DAP servers themselves.

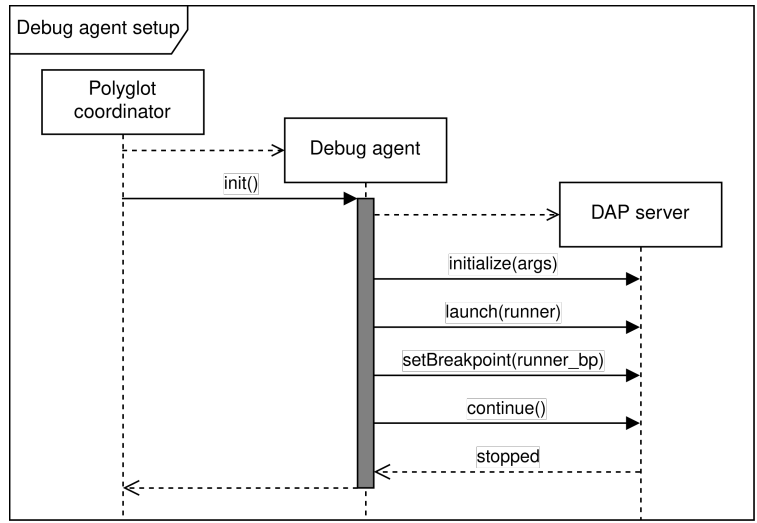

**Figure 3** Initialization phase of the debug agents in the polyglot debugger

Indeed, while starting a DAP server on a program poses no issues, complications arise when trying to simulate polyglot call behaviours. First, a DAP server finishing the execution of a program after a polyglot call would usually terminate and require a restart for each subsequent polyglot call to that language. Even without considering potential performance implications, this makes it impossible to have state preservation between different polyglot calls to the same language, which is a feature present in some polyglot platforms.

Furthermore, we need to be able to pause and resume execution of code, but also to be able to execute code while paused. Indeed, while the DAP server for language A is paused after a call to language B, it is possible that language B will call some code from language A again. Still for the same state preservation (and performance) concerns, we do not want to be forced to start another DAP server instance; the language A DAP server should execute the code while still being able to resume the original execution later on, when the call to language B finishes. In a concurrent program, this pausing interaction can happen on a per-thread basis as discussed further in Section 4.

Lastly, we reuse DAP servers as-is, without changing their implementation. This means that they are not syntactically aware of constructs introduced exclusively by the polyglot platform that is being emulated. In our example of simplified GraalVM behaviour, this means that the `polyglotEval` function is not defined by default in the scope of the DAP. Thus, if the DAP tries to execute code written for the platform containing a call to that function, it will throw an error and stop execution.

To resolve all of these problems, we introduce the polyglot runner.

## 3.3 Polyglot Runner

A polyglot runner is a program tied to a debug agent, written in the language targeted by the debug agent. The polyglot runner's role is to define the construct used for

■ **Listing 1** Pseudo-code of a typical polyglot runner defining the polyglotEval function and code to handle dynamic execution of incoming programs

```
1  def polyglotEval(language, foreignCode) { # the arguments are unused in the runner, but are
   ↪   retrieved by the agent for the outgoing execution request
2  🔴 ret = "foreign polyglot call result"
3     inputCode = "path/to/new/code"
4     while (inputCode != "") {
5        res = execute(input_code)
6  🔵 inputCode = "path/to/next/code"
7     } # while waiting on the outgoing request, we handle incoming execution requests
8     return ret
9  }
10
11 while (True) {
12 🔵 inputCode = "path/to/code" # inputCode is manually set to the next file by the agent
13    res = execute(inputCode)
14 }
```

polyglot calls in the scope of the DAP server, as well as provide a way to manage execution flow through smart breakpoint usage.

The way we can use the runner is by making it the actual program being executed by the DAP server. Rather than having a DAP server directly execute input code, it will execute the runner and that runner itself will execute the input code. This is why in Figure 3 we launch the runner rather than a program given by the polyglot debugger client. The debug agent will accordingly manage the runner and the DAP server to ensure the presence of the runner is transparent to the polyglot coordinator, making it appear as though only the input program is being executed.

Listing 1 provides pseudo-code of what a typical runner looks like when trying to instantiate our framework to GraalVM-like syntax and semantics.

The first task of the runner is to ensure the DAP server can parse and step over polyglot calls without errors. It thus defines the construct used to invoke polyglot calls in the source code; in our case and in Listing 1 this is the polyglotEval function that emulates a simplified version of GraalVM's polyglot.eval.

To avoid triggering undefined errors, this function needs to be visible in the scope of the input programs provided by the client. While there may be exceptions depending on the target language of the runner, most languages have a concept of a global scope which can be used for this purpose.

As the individual DAP servers are not able to execute the real implementation of polyglot calls, we need to ensure whenever a call is made to this polyglotEval function, execution is stopped so we can emulate the behaviour. To this end, we set a breakpoint at the start of the body of the function. We refer to this breakpoint as a polyglot breakpoint; every time the DAP server is stopped at this specific line, we know that it is because a polyglot call was invoked in the debug session. The debug agent can check for this, retrieve the arguments passed to the function, and notify the polyglot coordinator to continue execution through another DAP server. The polyglotEval function also returns a dummy ret variable. This variable's value can be changed according to the result of the foreign code execution, using the standard DAP setVariable request. Section 4 details more how this data can be converted to fit the type systems of each language involved.

As mentioned previously, it should be possible for a debug agent to start, pause and resume execution of an input program. While this first input program is paused, it should also be possible for the debug agent to receive another input program to be executed, and resume execution of the first input program later on.

The base execution loop for an agent is the loop at lines 11-14. As some input code is received by the debug agent, it sets the corresponding variable in the runner and then sends a continue request to the DAP server. This triggers the execute function at line 13 in Listing 1, which is a language-dependent mechanism to execute code. Once execution of the input program is over, the DAP server hits another breakpoint which we call a standby breakpoint, set at line 12. At this stage, the result of the execution is available in the res variable, and the debug agent is ready to accept a new input from the coordinator. This loop iterates until the entire polyglot debugging session is over. Figure 4 illustrates this idea through an example featuring two debug agents handling a continue request. Breakpoints present in the runner are highlighted by color in the figure.

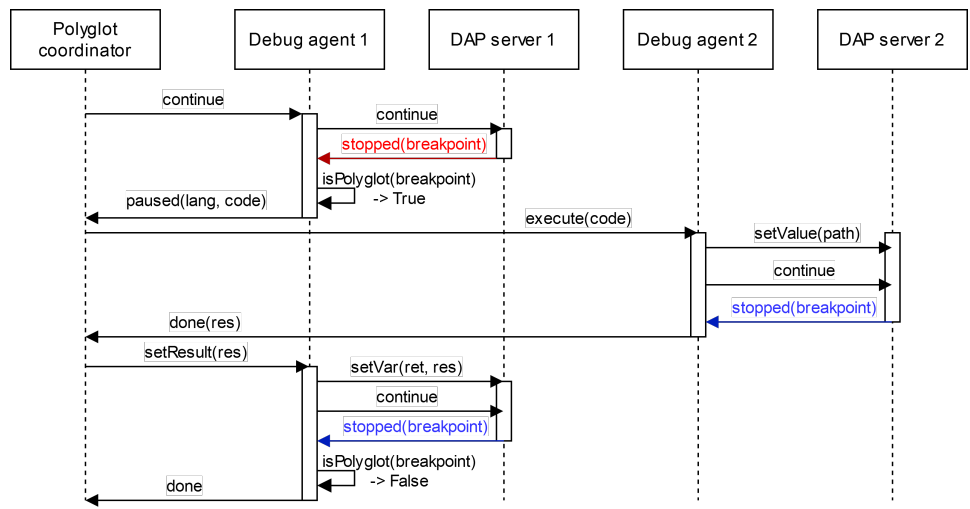

**Figure 4** Example of a continue request being handled by two agents

During the execution of an input program in language A, a polyglot breakpoint may be reached, pausing the execution due to a polyglot call to language B. However, language B may also invoke polyglot calls back towards language A, requiring the debug agent to stay available and await input programs. Spawning a new DAP server process for every incoming input program may lead to an exploding memory footprint, and hence, is not an ideal solution. Instead, the inside of the polyglotEval function contains another execution loop, seen at lines 4-7 in Listing 1. This secondary execution loop functions the same as the main one and also features a standby breakpoint at line 6, but remains available even while execution is paused and awaiting a result from the polyglot call. When the debug agent needs the runner to exit this loop, it empties the `input_code` variable in much the same way it usually sets its value. An example of this call-back interaction is show on Figure 5 where an agent receives a request to execute code while being paused after an outgoing request for foreign code execution. Relevant runner breakpoints are once again highlighted by color.

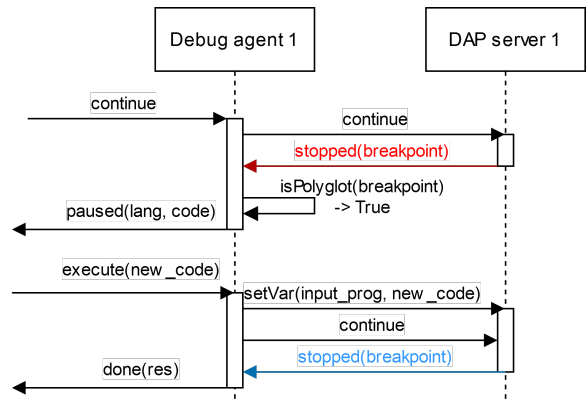

**Figure 5** Example of a debug agent handling an incoming polyglot request while being paused for an outgoing polyglot request

## 4 Framework Execution Model

While our framework aims to provide debugging capabilities agnostically of the targeted polyglot execution platform, it in fact exists as a polyglot platform of its own. We here discuss the execution model of this platform, and in particular some alternatives, trade-offs, and limitations of the framework.

### 4.1 Control Flow

Our handling of polyglot control flow operations is accomplished by programmatically setting under the hood a breakpoint at the line where a polyglot call occurs. The caller debug agent and polyglot coordinator then redirect polyglot debug requests to the callee debug agent required for the target language of the polyglot call. This interaction occurs recursively according to the nesting depth of the polyglot calls.

In our implementation of `polyglotEval`, polyglot calls are synchronous. Thus, the caller DAP server pausing during the execution of the callee is the expected behaviour. However this is not a restriction of the framework itself, as runners can implement polyglot operations in an asynchronous way, using the language's concurrency mechanisms. The caller DAP server would then have two threads: one continuing execution, and one awaiting for a response from the callee DAP server. As each DAP server is its own process at the operating system level, the only last thing to handle to support concurrency is to ensure the coordinator memorizes pairs of caller-callee threads to be able to redirect execution when the callee finishes. DAP provides a thread identifier in the response to stack requests, which can easily be leveraged for this.

As polyglotEval takes the form of a function call, the coordinator for our implementation logically implements a simple call stack mechanism initiated by requests to the agents to execute some code. However it is worth noting that DAP supports not just requests from the client to the server, but also reverse requests from the server to the client. It would thus be possible to support execution models that are more event-driven, where the agents initiate operations rather than the coordinator itself.

### 4.2 Data Exchange

While data flow and data exchange operations are not the main focus of the paper, they are nonetheless an important part of polyglot interactions. The main concern is how to convert data from the original language's type to the receiving language's type system when it is transferred.

In the context of the Debug Adapter Protocol, all messages are sent as strings. This means that each DAP server implements a canonical string representation for the values in the target language. As such, when transferring data, the sender debug agents can implement a conversion from the DAP representation to a common representation, while the receiving debug agents implement the reverse conversion for their own language. This concept of a common representation is how many polyglot platforms implement data conversion (*e.g.* GraalVM or WebAssembly).

Another challenge when handling polyglot data is the question of how to handle values by reference. In the general case this can be hard due to different representations for data across languages; for our architecture, DAP servers are their own separate processes with no shared memory. Thus it is necessary to emulate reference behaviour. This can be done by using data breakpoints, another feature of DAP which stops execution every time some data is accessed. This lets each DAP server hold their own copy of a shared object, and propagate potential changes made to it to other language by notifying the coordinator of a change. However, this implies a potentially very high amount of requests being passed through a string-based protocol, which could imply poor performance.

### 4.3 Matching Platform Semantics

One limitation to discuss is the semantic match between the execution platform and the instance of our framework. As our approach is agnostic of execution platforms, any of them can theoretically be supported. However, this is not an automated process: each time languages or a platform needs to be supported, an implementation has to be written to specialize a coordinator and agents correspondingly. This manual process can lead to errors which would make the debugger behave incorrectly.

However, we argue this is not a very error-prone process. Implementing the behaviour of each language is left up to the DAP server implementation. In the author's experience, while polyglot platforms are wonders of optimization, they also tend to not exhibit overly complex behaviours in the majority of use-cases. Even for the edge cases, the existence of a reference implementation in the polyglot platform itself greatly helps in designing tests for a polyglot debugger instance. Nonetheless, a human-driven process implies potential human errors, and this is an inherent limitation of our approach. An interesting future work perspective could be to study the generation of debuggers automatically based on a given polyglot platform.

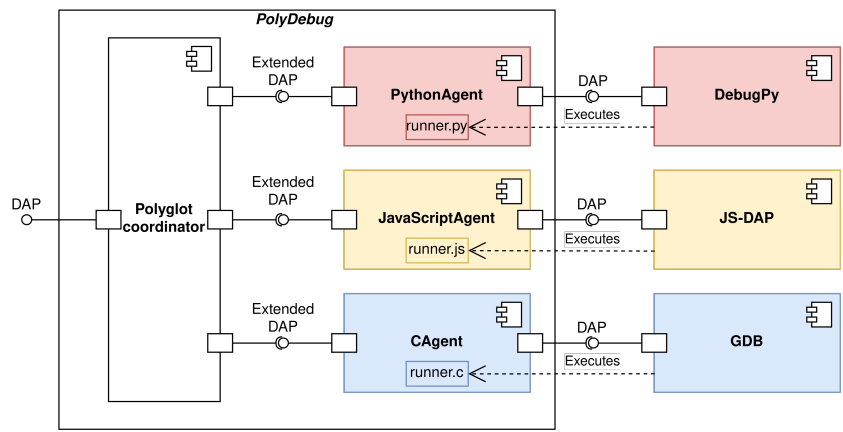

**Figure 6** The PolyDebug prototype

## 5 *PolyDebug* Implementation

This section describes our implementation of the approach. We first give an overview of some technical choices and the structure of the polyglot coordinator, and then we detail each individual debug agent and its specificities.

### 5.1 Overview and Polyglot Coordinator

Figure 6 shows the instance of our framework corresponding to our prototype. Our implementation of the approach is written in Python, running on version 3.12.4. For any pre-processing of the files given as input, we use the Tree-sitter[8] framework. To manage the various DAP servers, we use the `subprocess.Popen` function as well as Debugpy's[9] `JsonIOStream` module. Since these are used for communication with all the DAP servers, they allow us to create a debug agent interface for all debug agents that exposes our extended DAP. This interface also provides default implementations of many base DAP requests. Debug agents can then simply inherit from this interface and override the few requests they may need to customize.

The polyglot coordinator itself is implemented as a generic manager of a list of debug agents. It has no direct reference to any of the individual debug agents; debug agents are instead expected to be passed as arguments of the coordinator, which adds them to a list of managed debug agents.

---

[8] https://tree-sitter.github.io/tree-sitter/ (Visited on 2025-01-21).
[9] https://github.com/microsoft/debugpy/ (Visited on 2025-01-21).

The main polyglot interaction mechanism we implement is the execution of code written in another language, which can return a corresponding value. We call this a polyglot call, and follow a syntax similar to GraalVM with the `polyglotEval(language, code)` function. Both the syntax and the exact semantics of this polyglot interaction mechanism could be adapted to support various kinds of execution platforms.

When a request to run code is received, the coordinator automatically detects the appropriate debug agent based on the file extension. This debug agent becomes the active agent. From this point on, any incoming requests are redirected to the active agent. Each time a request completes, the polyglot coordinator checks the state of the debug agent. If a polyglot breakpoint has been reached, the coordinator keeps the debug agent stopped and switches the active agent to continue execution in the new language. The former active agent is added to a call stack. Once the active agent finishes execution, the call stack is used to go back to the previous debug agent. This process continues until the entire execution finishes.

## 5.2 Debug Agents and Runners

While the polyglot coordinator itself is agnostic of languages it handles, this is not the case for the debug agents. Each debug agent needs to manage the specificities of both the targeted language and the capabilities of the DAP server. We cover each of the respective languages implemented by our prototype, first discussing the structure of the runner and the mechanisms for code execution. We then describe specifics of the debug agent itself.

### 5.2.1 Python

The first debug agent targets Python, a dynamically typed, interpreted language. Python is a language renowned for its versatile standard library that allows to perform many operations at a high level of abstraction. Most notably, among its built-in functions we can find the `exec` and `eval` functions as well as the `ast` module, which allow to easily manipulate and execute Python programs. This exactly fits the needs of the runner, and thus the Python implementation of the runner is quite simple.

When receiving code to execute, the Python runner first uses the `ast` module from the standard library to parse the incoming code and split the last statement from the rest of the code. We then call the `exec` function on the first piece of code, and finally `eval` the last statement to get a return value. Since these functions allow to pass dictionaries to provide an execution context, we can simply provide a dictionary with the `polyglotEval` function to inject it into the executed code.

The debug agent itself does not have any major particularities linked to Python. It handles the runner execution flow and sets variables as expected, depending on the requests received from the polyglot coordinator. We chose to set the breakpoints in the runner traditionally through requests to the DAP server, but it is worth noting this could also be achieved by calling the builtin Python `breakpoint` function directly in the runner.

### 5.2.2 JavaScript

The next debug agent targets JavaScript, or more accurately Node.js which is a dynamically typed, interpreted language. JavaScript's standard library is less furnished than Python's in terms of program execution capabilities. Nonetheless, we can make use of the require built-in function. This function is normally used to load modules, but it also has the effect of executing the code contained in the file given as argument. However, it does require a few workarounds.

As require's intended purpose is loading modules, it stands to reason that the return value corresponds to the module.exports object of the target file. However, this object is by default empty. Thus, a statement to define the exports is needed at the end of the target file. In our implementation, we arbitrarily decided that defining these exports is the programmer's burden, as it is part of the polyglot semantics. However, alternatives are possible: similarly to Python, we could have determined that the last statement was the return value. In this case, a post-processing phase would have been needed to insert "module.exports =" in front of the last statement of the program.

Another problem is the require cache. This is a caching mechanism which aims to avoid reloading a module that was already loaded once. The problem is that this also skips execution of the code if the file was previously cached. This does not fit our purpose, since there is no reason that a given file could not be the subject of multiple polyglot calls in one execution. Thus we have to be mindful and, before executing a file, we need to clear the cache entry for this file. It is worth noting that this merely allows the code in that file to be re-executed, without clearing away any of the VM memory state it may have set up previously.

As for the definition of the polyglotEval function, there are no particular issues. For the execution loop, we need to handle the same two problems linked to our usage of require. Sharing the function with the executed code is simply a matter of defining the function as part of the global object. This does result in the debug agent having to check for global.polyglotEval on a polyglot breakpoint, rather than just polyglotEval. If we were using a JavaScript web browser runtime over a Node.js one, the agent would have to check for window.polyglotEval.

### 5.2.3 C

Unlike the previous two languages, C is a compiled and statically typed language. Additionally, C is not reputed for having a very large standard library. However, it benefits from its DAP server being implemented with GDB,[10] which supports a high variety of debugging features. This and a few other factors make the runner and debug agent quite atypical compared to the two previous ones.

First, the runner's main execution loop is completely empty. It only serves to keep the DAP server running, but has no actual code within the loop. The runner itself defines two functions, which are simply small wrappers around the dlopen and dlclose functions. These functions allow to dynamically load (and unload) compiled libraries into the current memory space of the program.

---

[10] https://sourceware.org/gdb/ (Visited on 2025-01-21).

The polyglotEval function is not defined in the runner itself, but in a separate polyglot_eval.c file. This is to simplify the injection process through include directives. The function also does not include an execution loop, unlike other runners. It is worth noting that in our implementation, we decided to implement polyglotEval as a function returning a double to simplify semantics. In a realistic implementation, this should instead either return a void* to handle any kind of return value, or be split into multiple functions for each type (*e.g.* polyglotEval_double, polyglotEval_char, *etc.*). This is left for future work improvement.

To account for this, we do need to pre-process source files, by adding a line to include the function at the start of called programs. This implies that the debug agent has to compile the code into a shared library object as part of the code execution process. Along this, we implement a simple caching mechanism to only compile each file once. After the code is compiled, the debug agent instructs the DAP server to call dlclose to unload any previous code, and then dlopen to load the new code.

Once this code is loaded, the debug agent takes over. Despite not having any code in the runner to execute code, the DAP server allows execution of GDB commands through the evaluate request. This allows us to access the very powerful capabilities of GDB, and most notably its call command. This command allows to perform a call to any function currently loaded in the program memory space. Thus, it effectively implements the code execution mechanism without the need for any supporting code in the runner. We can use the objcopy –redefine-sym command to rename the symbol of the main function in the code that we load (to avoid conflicting with the runner's main function), and then call it with GDB. After the function call finishes, the return value is contained in the $1 variable, allowing the debug agent to easily retrieve it.

### 5.3 Visual Studio Code Extension

As a showcase of the prototype, we have implemented a Visual Studio Code debug extension based on the re-exposed DAP interface, available online.[11] This implementation reuses the VSCode's documentation's template debug extension,[12] with only the bare minimum of naive changes to plug our prototype in place of the default debug server. As this template's API is compliant with DAP, it also serves as a test that we properly re-expose the protocol: all debug requests are simply redirected directly to the prototype with basically no extra work on the IDE extension side.

As we can see in Figure 7, most features of VSCode's debug view function as intended, allowing us to step through the program while hitting various breakpoints across languages. The main point of interest is the stack and variables view, which includes the runner's values, and only presents the current language's view. However these are merely technical choices; the runner is always the first level of the stack and can thus easily be filtered out, while the coordinator holds a representation of a call stack for each language and could thus reconstruct the entire stack across languages.

---

[11] https://github.com/phoudail/polydebug_demo (Visited on 2025-01-21).
[12] https://github.com/microsoft/vscode-mock-debug (Visited on 2025-01-21).

**PolyDebug: A Framework for Polyglot Debugging**

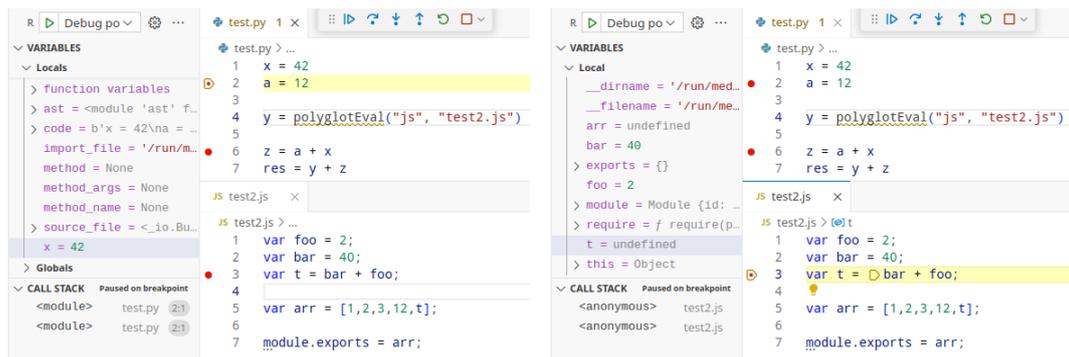

**(a)** The first breakpoint is hit in test.py.

**(b)** The view automatically switches to test2.js once execution flows inside the polyglot call.

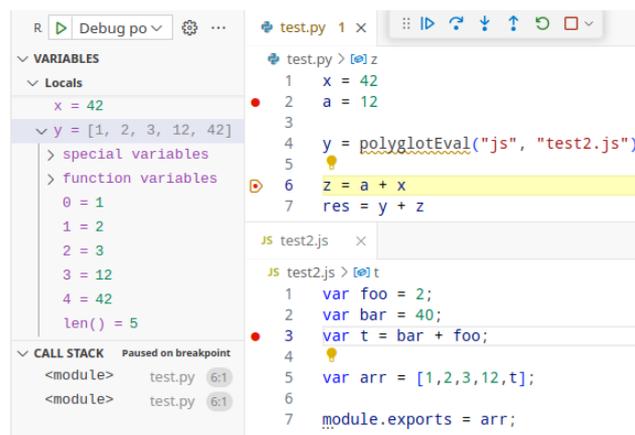

**(c)** Back to Python after the Javascript finishes; we can see in the stack view y now holds the value of arr.

■ **Figure 7** A debug session on a toy program in our VSCode extension

## 6 Evaluation

We claim our architecture to be extensible and language-agnostic. While the language-agnosticism is by design of the polyglot coordinator, this section aims to verify extensibility by investigating the development effort required to implement and extend our prototype. Additionally, we also evaluate the performance of our prototype to ensure our architecture results in a usable tool.

### 6.1 Research Questions

In order to assess the extensibility and performance of our approach, we formulate the following research questions:

- **RQ1** → *What is the implementation effort required to build a polyglot debugger following this architecture?* This aims to investigate our extensibility claim, by looking at the effort required in our own implementation and assessing limiting factors.

■ **Table 1** SLOC spread for our implementation of the architecture

| Component | SLOC | |
|---|---|---|
| Polyglot coordinator | 118 | 362 generic SLOC |
| Debug agent interface | 244 | |
| Python (agent + runner) | 147 + 25 | 172 to 259 language-specific SLOC |
| JavaScript (agent + runner) | 221 + 38 | |
| C (agent + runner + polyglot lib) | 231 + 14 + 4 | |
| **Total** | 1042 | |

- **RQ2** → *What is the overhead introduced by our approach?* This aims to measure the added delay that our approach might add on top of the delay already introduced by being in a debug context. More specifically, this can be refined into two sub-questions:
  - **RQ2.1** → *How does our approach impact the execution time of polyglot programs?* This aims to verify that given a polyglot program, our approach is able to execute it with sufficient performance to allow for debug activities to occur smoothly.
  - **RQ2.2** → *How does our approach scale with the number of polyglot calls?* This aims to check to which extent our approach is able to handle more extreme cases of polyglot programming, with very frequent calls between languages.

## 6.2 RQ1 (Implementation Effort)

In this section, we answer RQ1 by discussing our own implementation of the approach. In total, our prototype represents 1042 Source Lines of Code (SLOC) measured by the cloc [2] tool and shown in Table 1. This is spread across seven classes and nine files, four of which are runner files (three language runners plus the C `polyglot_eval` file).

Overall, the amount of SLOC in our implementation is fairly low for the final achieved polyglot debugging capability. While we do not support the entirety of possible debugging operations, we do provide support for the basic ones: stepping, setting breakpoints, going to the next breakpoint, and examining the current scope with its variables. All of these operations are implemented by the polyglot coordinator in a completely language-agnostic way, relying only on the generic debug agent interface and request redirection. Most of the debug agents did not have to modify the default behaviour of the DAP requests corresponding to these operations, meaning other debugging operations could likely be supported in a similarly concise way provided the DAP server provides the adequate capabilities.

Adding support for a new language is as simple as adding a new debug agent and its corresponding runner; the polyglot coordinator and the agent interface can be reused as is. Table 1 showcases a range between 172 and 259 SLOC for the combination of a debug agent and its runner. It is difficult to estimate how many SLOC a new debug agent and runner would require, as this could potentially highly vary depending on the targeted language. Nonetheless, we can estimate that support for a new language would remain in the same order of magnitude in terms of implementation efforts – meaning informally, it would require a few hundred SLOC. This seems to still suggest a reasonably low implementation effort.

As reference, and while it is not very relevant to directly compare the two approaches, the GraalVM debugger instrument stands at around 300 lines of code; the GraalJS and GraalPython Truffle languages repositories stand at respectively 12 million and 2.8 million lines of code while the DAP servers for C, JavaScript and Python average 150 thousand lines of code as measured by the SCC[13] tool. It is worth noting that the Graal projects are full language implementations including optimization code, which is impossible to decouple from debugging code due to the nature of the Truffle framework. While SLoC are by no means an absolute or objective metric to measure engineering complexity, this still seems to suggest that our approach does not require such a high implementation effort. However, the arguably most difficult parts of the implementation effort are not reflected in the SLoC count.

First, while a debug agent does not need any knowledge of the other languages in the system, it is not agnostic to itself. As showcased in Section 5.2, both the agent and the runner can vary in their implementation depending on the targeted language. This requires some knowledge of language mechanisms allowing code execution.

Additionally, implementing the debug agent can also require knowledge of the debugger of the target language. While DAP provides a standard interface to communicate with debuggers, it also leaves room for debugger-specific implementation choices. As a result, the DAP specification does not specify arguments for many requests, and this instead varies among DAP servers. Furthermore, some requests do not behave in the same fashion even if they have the same arguments. With the Python DAP, the `evaluate` request can evaluate any expression and respond with the evaluation result. With the C DAP, the same request can evaluate a variable name, but also provide access to the GDB command-line interface if the request content is prefaced with `-exec`. These differences can make implementing a debug agent tricky without prior knowledge of the DAP server, in a way not reflected in the SLOC count.

One example of language to which we could not successfully apply our approach is Go. This was due to two main factors: first, the Go DAP Server[14] had some restrictions on supported language features (such as the `interface{}` type in function parameters) which limited our ability to execute arbitrary incoming code in a debug context, and also did not support the `setExpression` request which could have served as a workaround. This highlights the importance of our assumptions from Section 2 in order for the approach to be applicable.

> $RQ_1$ **insights:** Overall, our approach requires a fairly low amount of development effort to implement, provided pre-existing DAP servers. However, there are still some implementation challenges related to the expertise of handling a debugger through DAP as well as individual language mastery.

---

[13] https://github.com/boyter/scc (Visited on 2025-01-21).
[14] https://github.com/go-delve/delve (Visited on 2025-01-21).

## 6.3 RQ2 (Performance)

In order to answer RQ2, we conduct a performance evaluation on our implementation of the approach. We first describe the dataset and protocol used to conduct the experiments, and then present results along with a discussion of these results.

### 6.3.1 Dataset and Experimental Protocol

In order to build a set of polyglot benchmarks, we relied on the Benchmarks Game [8] which provides implementations of the same algorithms in a variety of programming languages. In fact, some of these implementations are transliterations of each other, meaning they share the same organization of code, *e.g.* calls to f1(), f2(), f3(), etc. This allows us to easily replace one part of a program with the same part of the same program written in another language, thus creating a polyglot version of the program. For example, we could call f1() in its original source language C and the f2() from the Python implementation – thus, making a semantically equivalent polyglot program to the original. We manually selected three programs of the benchmarks (namely *fasta*, *nbody*, and *pidigits*) which had transliterations in C, Python, and JavaScript, which gave us a base set of nine programs. The *fasta* benchmark computes DNA sequences, first by copying a given sequence and then through random selection. The *nbody* benchmark models the orbit trajectory of planets using a simple symplectic integrator. The *pidigits* benchmark generates digits of Pi through arbitrary precision arithmetic.

As a baseline, we executed each of these original non-polyglot programs with the DAP servers. Then, we created polyglot variants of the original benchmark programs, by replacing functions or parts of the algorithms by their counter-part in another language. For instance, the *fasta* program is divided in three parts, corresponding to the generation of three DNA sequences. In this program's case, we replace each of these parts with their implementation in another language, thoroughly producing both bilingual and trilingual variants covering each language combination.

Additionally to this set of benchmarks, we also built a simple program in each of the three languages to stress test our implementation with regards to the number of polyglot calls. This program takes an integer $n$ as parameter, and iterates $n$ times, with each iteration doing one polyglot call. We ran this iteration for multiple values of $n$ and for every language combination, in order to see how our implementation scales with the number of polyglot calls.

The experiments ran on a Dell Precision 3581 workstation, equipped with a 13th generation Intel Core i9-13900H CPU, 32 GB of RAM and running Manjaro 6.9.3-3 x86_64. We now present the results of these experiments and discuss them.

### 6.3.2 Results

Figure 8 shows the results of the first experiment of the set of benchmarks. In the figure, each blue line represents the execution time of one polyglot program, scaled as a percentage of the range between the fastest and slowest single-language version of the program, as measured in our baseline. For instance, if a given program finishes in 100 seconds in Python, in 50 seconds in JavaScript and 20 seconds in C, we scale the execution time of Python-JavaScript polyglot programs so that 100 seconds represents

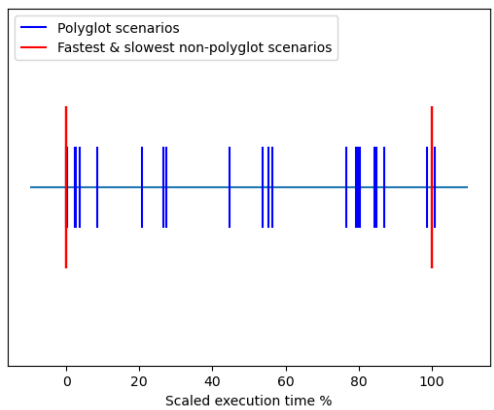

Figure 8 Benchmarks results, with data scaled as a percent of the slowest and fastest single-language runs

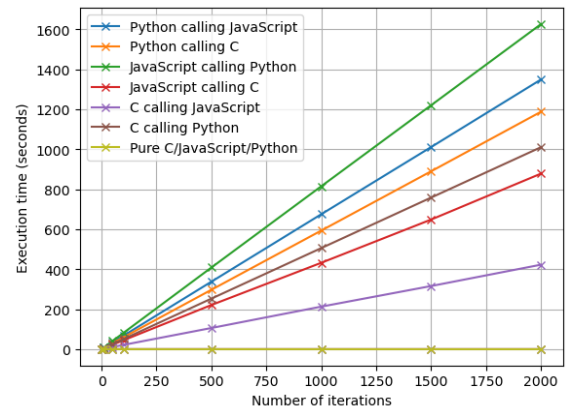

Figure 9 Execution time of the stress test depending on amount of loop iterations

100 % while 50 seconds represents 0% on our graph. For the Python-JavaScript-C polyglot program, 0 % is instead scaled to 20 seconds, the fastest single-language baseline. The baseline single-language programs are represented as red lines.

According to the results showcased by our set of benchmarks in Figure 8, it does not seem that our solution introduces significant overhead. Indeed, nearly all of the polyglot versions of the programs finish execution within the time range defined by the slowest and fastest execution. The only scenario where execution time was higher than the upper limit of our time range (on average 94.5 seconds to execute, versus 93.9 for the upper limit of our range) was when we replaced the entire code of a program with a single polyglot call to an alternate implementation. Variance within this time range can otherwise be explained by the difference in which languages are called for each polyglot program, and the respective performance of the DAP servers for those languages. While it is not the purpose of our evaluation to examine this, the fastest DAP server in these benchmarks was the one for C, while the slowest one was Python; this did vary, however, depending on the benchmark program.

The results of the second experiment, *i.e.* the loop stress test, are shown in Figure 9. The graph shows the relation between the amount of iterations and the execution time, in seconds. Each individual line corresponds to a combination of calling and called language, *e.g.* the purple line corresponds to a C loop making a polyglot call to JavaScript on every iteration. We additionally show measures for the loops without any polyglot calls in every language, which appear flat due to the very low execution times compared to the polyglot versions.

Despite promising earlier results, the stress test illustrated in Figure 9 shows that our solution does introduce some overhead. More precisely, it introduces between 0.2 and 0.8 seconds of overhead per polyglot call during execution. This amount varies depending on language combinations. The execution times in our set of benchmarks from the first experiment varied between 12 and 1600 seconds, and with a maximum of 4 polyglot calls per execution; thus, it was not impacted by this overhead.

> *RQ$_2$* **insights:** Our results show that our solution does not introduce significant overhead with regards to the execution of each individual sub-program of the polyglot program being debugged. However, it does introduce overhead through the polyglot calls allowing these sub-programs to interact. This overhead scales with the number of polyglot calls during program execution.

### 6.3.3 Discussion

Overall, the results of our experiment show potential for a very high runtime overhead, dependent of the number of polyglot calls at runtime. However, our approach is not in the context of execution, but of debugging for polyglot programs.

Debugging is an interactive task, with regular input from the user. In this context, a program does not necessarily need to execute as fast as it would under normal circumstances, since there is a human in the loop. So long as tasks such as stepping do not exceed acceptable latency thresholds, some degree of overhead is acceptable. However, there are still problems that may be caused by excessive amounts of polyglot calls. For instance, a debugging session might use breakpoints to examine a late point in the program's execution after a large number of polyglot calls. If our approach introduces significant delay in reaching this breakpoint, it impairs the ability to efficiently debug the program.

The overhead introduced in our approach scales linearly with the number of polyglot calls: the more polyglot calls there are, the more overhead there is. More accurately, this overhead would only be significantly perceived by a human performing debugging activities if the sub-program being called has an execution time in the same order of magnitude as our overhead does. However, we argue this is unlikely to be the case.

Polyglot programming is a practice that aims to use the language best suited for solving a given problem. Requiring the usage of another language implies that the current language is sufficiently unpractical to solve the problem with. With how versatile modern programming languages have become, this suggests that polyglot calls will be towards relatively large sub-programs solving non-trivial problems. While further study of real-world polyglot programming practices is required to confirm this, we argue that these sub-programs would thus have execution times significantly higher than our approach's overhead, keeping debug sessions responsive from the user's point of view.

Despite this, there may still be cases where a polyglot program is structured in such a way that there are many polyglot calls to sub-programs with small execution times. In those cases, our approach is not a good fit, and other solutions should be investigated to perform debugging activities.

### 6.4 Threats to Validity

This section discusses threats to the internal and external validity of our conclusions drawn in the previous sections.

### 6.4.1 Internal Validity

A first threat to the internal validity of our results is the accuracy of our measurements. When measuring the execution time of a program, there is often the possibility of noise occurring, which can impact the representativity of obtained results. To mitigate this threat, we follow a rigorous experimentation protocol, where each measure is done in the same execution conditions and with our prototype being the only program running on the test machine. Additionally, as part of the protocol we execute each individual scenario 10 times and compute the mean to get our result data. We also ensure that all our scenarios take at least a few seconds to complete, so that our findings are less impacted by small variations in execution times.

Another threat is the qualitative nature of our discussion on implementation effort. As this can be a subjective factor, there is the possibility of our discourse being biased. To try and mitigate this, we use the SLOC metric in an attempt to assess the implementation effort more accurately. Nonetheless, there is more nuance than simply program size. Thus, we also attempt in our discussion to identify various limiting factors not necessarily represented with SLOC count.

### 6.4.2 External Validity

Some external validity threats lie in the way we build our experiment dataset. Since we used "toy programs" rather than real-world application to measure performance, there is a risk that our findings cannot be generalized to real-world software. Additionally, we created the programs in the dataset ourselves, potentially introducing bias in the way programs are structured that influences our findings. We mitigate these factors by basing ourselves off of a benchmark set already used by other research work for performance measurements. We also modified the original programs as little as possible, meaning aside from glue polyglot calls none of the source code originated from us nor was modified by us. Still, there are potential issues in how representative of real-world programs our dataset is, which we discussed previously in Section 6.3.3. Further experiments on other datasets remain necessary before to attempt to generalize our results.

Another threat lies in our set of implemented languages. As we only provide support for Python, JavaScript, and C, it is difficult to fully assert that our approach applies to other languages that exist. Nonetheless, we cover both dynamically typed, interpreted languages and a statically typed, compiled language. Additionally, we rely mostly on mechanisms for function calls and modules, which most (if not all) general purpose programming languages implement. Thus, we are confident that our approach fits the vast majority of these languages. To further verify this, part of our future work consists in implementing support for more languages and attempting to automate the process of supporting new languages.

## 7 Related Work

Polyglot debugging has not received much attention from the programming languages and compiler construction community. Strategies for polyglot (or multi-language) debugging can be divided into two main categories. The first involves mechanisms for language interoperability, such as foreign functions interfaces, coordination architectures, or dedicated protocols. The second presumes that multiple languages execute in the same runtime environment. Below we discuss relevant work in both categories.

Debugging heterogeneous systems (another term for polyglot software) was pioneered by Olivier et al. in the context of the TIDE debugger architecture [23, 24, 29]. Different runtime environments for different languages send relevant debugging events to a central coordination architecture (called the TOOLBUS), which aggregates and passes on the events to the debugger client components for inspection by the programmer. This approach is similar to our work, except that we rely on a standard protocol (DAP), thereby restricting the design space to language runtimes that fit the constraints of the protocol. The TIDE approach offers more flexibility at the cost of more custom engineering.

A similar system for multi-language debugging is Blink [14, 15, 16]. In this case the transition between languages and their runtimes is mediated through FFIs. The system demonstrates multi-language stack traces and source location tracking. Nevertheless, in this case, the composition needs to be implemented for each language combination, because FFIs cannot be standardized.

Niephaus et al. [21] describe an approach to proxy foreign language runtimes into the Smalltalk message sending model; this allows them to adapt Smalltalk tools like the debugger to become multi-language. However, their approach is different from the FFI strategy, because FFIs require going through the operating system, which yields control from the Smalltalk processes. The solution they propose is a multi-language VM based on RPython. So in a sense, this is similar to the .NET common language runtime (CLR), which has been used for multi-language debugging as well [27].

In the context of domain-specific language (DSL) engineering, Pavletic et al. [25] discuss extensible debuggers, i.e., given a base language with a debugger, how to modularly extend the debugger when the language is extended with new constructs. In a sense this is a special case of polyglot debugging. However, since the topic is language *extension*, the approach presumes that there is only a single runtime environment.

Bousse et al. [1] also discuss how to provide omniscient debugging in a generic manner for DSLs. Their solution allows to provide omniscient debugging support automatically for any new DSL through a common set of debugging facilities and adaptations to specific execution engines. However, the approach does not take into account any potential coordination between the languages.

Closest to our approach to debugger composition is the work on domain-specific debug protocols by Enet et al. [5, 6]. In this work, the notion of debug protocol is made explicit but tailored to DSLs and modeling languages. It is not necessarily aimed at composing debuggers, but leverages the same kind of abstractions to ease the development of debuggers for new languages and DSLs. The polyglot debugging of multiple such DSLs presents an interesting avenue for further research.

Vraný and Píše present a generic design pattern to compose debuggers of different languages [30]. Again, the idea is similar to the approach of this paper, with the difference that the framework is not standardized, and requires a single execution environment (basically interpreters implemented in Smalltalk).

Language extensibility for non-debugger tools can also be considered related to our approach. Many non-polyglot, but multi-language tools (*i.e.* with support for multiple individual languages in the same tool) adopt language-agnostic approaches to build a common baseline of features, which can be further specialized with language-specific inputs. Examples include Jupyter[15] notebooks which can swap kernels to support different languages, or the Tree-sitter framework[16] that can generate parsers for different languages by being provided with a grammar.

For polyglot tooling, the most well-know approach to genericity is GraalVM's Truffle framework [28, 31], which serves both as a language implementation framework and an instrumentation tool framework. The major advantage provided by Truffle tools is that they can be defined completely agnostic of languages thanks to the Truffle node and node annotation system: this means that a tool is defined once and then available for all present and future languages supported within Truffle. The drawback is that Graal and Truffle constitute a self-contained ecosystem: this means that for a language to be supported by Truffle instruments, that language needs to be (re)implemented using Truffle, without being able to reuse existing implementations and tooling.

To the best of our knowledge, our approach is the first to propose a novel extensible and language-agnostic framework for polyglot debugging based on reuse of DAP debuggers.

## 8    Conclusion and Perspectives

In this paper, we present a novel approach to building debuggers, which provides extensibility and polyglot programming support through a language-agnostic architecture while promoting reuse of existing language-specific debuggers. We show the applicability of our framework through a prototype implementation, *PolyDebug*. We further validate our approach through both an analysis on the effort of implementation required for our prototype, as well as the overhead it introduced which we deemed to be acceptable in the context of interactive debugging activities. However, there are still scenarios where our approach is not suited, due to the overhead scaling based on the number of polyglot calls at program runtime.

Future work will focus on many aspects of the approach, such as optimization mechanisms to improve performance; both caching and pipelining mechanisms could be explored to reduce the performance overhead. Another potential direction is in further improving extensibility of the approach, such as by providing a level of automation for support of new languages.

**Acknowledgements**    The research leading to these results has received funding from the *ANR* agency under grant *ANR JCJC MC-EVO$^2$ 204687*.

---

[15] https://jupyter.org/ (Visited on 2025-01-21).
[16] https://tree-sitter.github.io/tree-sitter/ (Visited on 2025-01-21).

**References**

[1] Erwan Bousse, Dorian Leroy, Benoit Combemale, Manuel Wimmer, and Benoit Baudry. "Omniscient debugging for executable DSLs". In: *Journal of Systems and Software* 137 (2018), pages 261–288. DOI: 10.1016/j.jss.2017.11.025.

[2] Albert Danial. *cloc: v2.00*. Version v2.00. Feb. 2024. DOI: 10.5281/zenodo.10674528.

[3] Lukas Diekmann and Laurence Tratt. "Eco: A Language Composition Editor". In: *Software Language Engineering*. Edited by Benoît Combemale, David J. Pearce, Olivier Barais, and Jurgen J. Vinju. Cham: Springer International Publishing, 2014, pages 82–101. ISBN: 978-3-319-11245-9. DOI: 10.1007/978-3-319-11245-9_5.

[4] Jan Ehmueller, Alexander Riese, Hendrik Tjabben, Fabio Niephaus, and Robert Hirschfeld. "Polyglot Code Finder". In: *Conference Companion of the 4th International Conference on Art, Science, and Engineering of Programming*. <Programming> '20. Porto, Portugal: Association for Computing Machinery, 2020, pages 106–112. ISBN: 978-1-45-037507-8. DOI: 10.1145/3397537.3397559.

[5] Josselin Enet, Erwan Bousse, Massimo Tisi, and Gerson Sunyé. "On the Suitability of LSP and DAP for Domain-Specific Languages". In: *2023 ACM/IEEE International Conference on Model Driven Engineering Languages and Systems Companion (MODELS-C)*. 2023, pages 353–363. DOI: 10.1109/MODELS-C59198.2023.00066.

[6] Josselin Enet, Erwan Bousse, Massimo Tisi, and Gerson Sunyé. "Protocol-Based Interactive Debugging for Domain-Specific Languages". In: *Journal of Object Technology* 22.2 (July 2023). The 19th European Conference on Modelling Foundations and Applications (ECMFA 2023), 2:1–14. ISSN: 1660-1769. DOI: 10.5381/jot.2023.22.2.a6.

[7] Hans-Christian Fjeldberg. "Polyglot Programming". Master thesis. Norwegian University of Science and Technology, 2008. URL: http://theuntitledblog.com/wp-content/uploads/2008/08/polyglot_programming-a_business_perspective.pdf.

[8] Isaac Gouy. *The Computer Language Benchmarks Game.* https://benchmarksgame-team.pages.debian.net/benchmarksgame/. (Visited on 2025-01-21).

[9] Andreas Haas, Andreas Rossberg, Derek L. Schuff, Ben L. Titzer, Michael Holman, Dan Gohman, Luke Wagner, Alon Zakai, and JF Bastien. "Bringing the Web up to Speed with WebAssembly". In: *Proceedings of the 38th ACM SIGPLAN Conference on Programming Language Design and Implementation*. PLDI 2017. Barcelona, Spain: Association for Computing Machinery, 2017, pages 185–200. ISBN: 978-1-45-034988-8. DOI: 10.1145/3062341.3062363.

[10] Philémon Houdaille, Djamel Eddine Khelladi, Romain Briend, Robbert Jongeling, and Benoit Combemale. "Polyglot AST: Towards Enabling Polyglot Code Analysis". In: *2023 27th International Conference on Engineering of Complex Computer Systems (ICECCS)*. IEEE. 2023, pages 116–125. DOI: 10.1109/ICECCS59891.2023.00023.

[11] Philémon Houdaille, Djamel Eddine Khelladi, Benoit Combemale, and Gunter Mussbacher. "On Polyglot Program Testing". In: *Companion Proceedings of the 32nd ACM International Conference on the Foundations of Software Engineering (FSE Companion'24)*. 2024. DOI: 10.1145/3663529.3663787.

[12] Sebastian Kloibhofer, Thomas Pointhuber, Maximilian Heisinger, Hanspeter Mössenböck, Lukas Stadler, and David Leopoldseder. "SymJEx: Symbolic Execution on the GraalVM". In: *Proceedings of the 17th International Conference on Managed Programming Languages and Runtimes*. MPLR '20. Virtual, UK: Association for Computing Machinery, 2020, pages 63–72. ISBN: 978-1-45-038853-5. DOI: 10.1145/3426182.3426187.

[13] Jacob Kreindl, Daniele Bonetta, Lukas Stadler, David Leopoldseder, and Hanspeter Mössenböck. "Multi-Language Dynamic Taint Analysis in a Polyglot Virtual Machine". In: *Proceedings of the 17th International Conference on Managed Programming Languages and Runtimes*. MPLR '20. Virtual, UK: Association for Computing Machinery, 2020, pages 15–29. ISBN: 978-1-45-038853-5. DOI: 10.1145/3426182.3426184.

[14] Byeongcheol Lee. "Language and Tool Support for Multilingual Programs". PhD thesis. University of Texas at Austin, 2011. URL: http://hdl.handle.net/2152/ETD-UT-2011-08-4084.

[15] Byeongcheol Lee, Martin Hirzel, Robert Grimm, and Kathryn S. McKinley. "Debug all your code: portable mixed-environment debugging". In: *Proceedings of the 24th ACM SIGPLAN Conference on Object Oriented Programming Systems Languages and Applications*. OOPSLA '09. Orlando, Florida, USA: Association for Computing Machinery, 2009, pages 207–226. ISBN: 978-1-60-558766-0. DOI: 10.1145/1640089.1640105.

[16] Byeongcheol Lee, Martin Hirzel, Robert Grimm, and Kathryn S. McKinley. "Debugging mixed-environment programs with Blink". In: *Software: Practice and Experience* 45.9 (2015), pages 1277–1306. DOI: 10.1002/spe.2276.

[17] *Lisp FFI documentation*. https://lispcookbook.github.io/cl-cookbook/ffi.html. (Visited on 2025-01-21).

[18] Philip Mayer, Michael Kirsch, and Minh Anh Le. "On multi-language software development, cross-language links and accompanying tools: a survey of professional software developers". In: *Journal of Software Engineering Research and Development* 5 (2017), pages 1–33. DOI: 10.1186/s40411-017-0035-z.

[19] *MetaCall Official web page*. https://metacall.io/. (Visited on 2025-01-21).

[20] Gunter Mussbacher, Benoit Combemale, Jörg Kienzle, Lola Burgueño, Antonio Garcia-Dominguez, Jean-Marc Jézéquel, Gwendal Jouneaux, Djamel-Eddine Khelladi, Sébastien Mosser, Corinne Pulgar, Houari Sahraoui, Maximilian Schiedermeier, and Tijs van der Storm. "Polyglot Software Development: Wait, What?" In: *IEEE Software* 41.4 (2024), pages 124–133. DOI: 10.1109/MS.2023.3347875.

[21] Fabio Niephaus, Tim Felgentreff, Tobias Pape, Robert Hirschfeld, and Marcel Taeumel. "Live Multi-language Development and Runtime Environments". In: *The Art, Science, and Engineering of Programming* 2.3 (Mar. 29, 2018), 8:1–8:30. ISSN: 2473-7321. DOI: 10.22152/programming-journal.org/2018/2/8. arXiv: 1803.10200v1 [cs.PL].

[22] Fabio Niephaus, Patrick Rein, Jakob Edding, Jonas Hering, Bastian König, Kolya Opahle, Nico Scordialo, and Robert Hirschfeld. "Example-Based Live Programming for Everyone: Building Language-Agnostic Tools for Live Programming with LSP and GraalVM". In: Onward! 2020. Virtual, USA, 2020, pages 1–17. ISBN: 978-1-45-038178-9. DOI: 10.1145/3426428.3426919.

[23] Pieter A. Olivier. "A Framework for Debugging Heterogeneous Applications". PhD thesis. University of Amsterdam, 2000. URL: http://hdl.handle.net/11245/1.193204.

[24] Pieter A. Olivier. "Debugging Distributed Applications Using a Coordination Architecture". In: *Coordination Languages and Models, Second International Conference, COORDINATION '97, Berlin, Germany, September 1-3, 1997, Proceedings*. Edited by David Garlan and Daniel Le Métayer. Volume 1282. Lecture Notes in Computer Science. Springer, 1997, pages 98–114. DOI: 10.1007/3-540-63383-9_75.

[25] Domenik Pavletic, Markus Voelter, Syed Aoun Raza, Bernd Kolb, and Timo Kehrer. "Extensible Debugger Framework for Extensible Languages". In: *Reliable Software Technologies – Ada-Europe 2015*. Edited by Juan Antonio de la Puente and Tullio Vardanega. Cham: Springer International Publishing, 2015, pages 33–49. ISBN: 978-3-319-19584-1. DOI: 10.1007/978-3-319-19584-1_3.

[26] Alexander Riese, Fabio Niephaus, Tim Felgentreff, and Robert Hirschfeld. "User-Defined Interface Mappings for the GraalVM". In: *4th International Conf. on Art, Science, and Engineering of Programming*. <Programming> '20. Porto, Portugal, 2020, pages 19–22. ISBN: 978-1-45-037507-8. DOI: 10.1145/3397537.3399577.

[27] Dennis Strein and Hans Kratz. "Design and Implementation of a high-level multi-language .NET Debugger". In: *.NET Technologies 2005* (2005), page 57. URL: http://oot.zcu.cz/NET_2005/Papers/Full/A83-full.pdf (visited on 2025-01-21).

[28] Michael Van De Vanter, Chris Seaton, Michael Haupt, Christian Humer, and Thomas Würthinger. "Fast, Flexible, Polyglot Instrumentation Support for Debuggers and other Tools". In: *The Art, Science, and Engineering of Programming* 2.3 (Mar. 29, 2018). Publisher: AOSA, Inc., 14:1–14:30. ISSN: 2473-7321. DOI: 10.22152/programming-journal.org/2018/2/14.

[29] Mark G.J. Van Den Brand, Bas Cornelissen, Pieter A. Olivier, and Jurgen J. Vinju. "TIDE: A Generic Debugging Framework — Tool Demonstration —". In: *Electronic Notes in Theoretical Computer Science* 141.4 (2005). Proceedings of the Fifth Workshop on Language Descriptions, Tools, and Applications (LDTA 2005), pages 161–165. ISSN: 1571-0661. DOI: 10.1016/j.entcs.2005.02.056.

[30] Jan Vraný and Michal Píše. "Multilanguage Debugger Architecture". In: *SOFSEM 2010: Theory and Practice of Computer Science*. Edited by Jan van Leeuwen, Anca Muscholl, David Peleg, Jaroslav Pokorný, and Bernhard Rumpe. Berlin, Heidelberg: Springer Berlin Heidelberg, 2010, pages 731–742. DOI: 10.1007/978-3-642-11266-9_61.

[31] Christian Wimmer and Thomas Würthinger. "Truffle: A Self-Optimizing Runtime System". In: *Proceedings of the 3rd Annual Conference on Systems, Programming, and Applications: Software for Humanity*. SPLASH '12. Tucson, Arizona, USA: Association for Computing Machinery, 2012, pages 13–14. ISBN: 978-1-45-031563-0. DOI: 10.1145/2384716.2384723.

[32] Thomas Würthinger, Christian Wimmer, Andreas Wöß, Lukas Stadler, Gilles Duboscq, Christian Humer, Gregor Richards, Doug Simon, and Mario Wolczko. "One VM to Rule Them All". In: *Proc. of Onwards!'13*. 2013, pages 187–204. DOI: 10.1145/2509578.2509581.

## About the authors

**Philémon Houdaille** is a Ph.D. student at the IRISA lab in the DiverSE team, Université Rennes 1, 35000 Rennes, France. He is interested in programming languages, program analysis, and model-driven engineering approaches, and his Ph.D. research focuses on polyglot programming technologies and tools. Contact him at philemon.houdaille@irisa.fr and https://phoudail.github.io/.
https://orcid.org/0009-0005-3538-5275

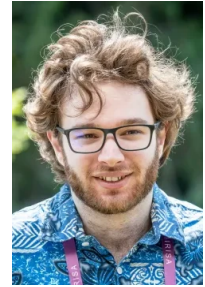

**Djamel Eddine Khelladi** is a CNRS researcher at the IRISA lab in the DiverSE team, Université Rennes 1, 35000 Rennes, France. His research interests are model-driven engineering, software engineering, software evolution and co-evolution, scaling code analysis, incremental build, and software processes. Khelladi received his Ph.D. in computer science from the University of Paris 6. Contact him at djamel-eddine.khelladi@irisa.fr and http://people.irisa.fr/Djamel-Eddine.Khelladi/.
https://orcid.org/0000-0002-2218-650X

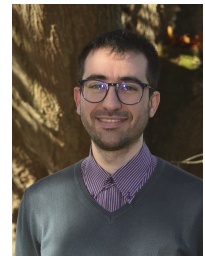

**Benoit Combemale** is a full professor of software engineering at the ESIR, University of Rennes, 35065 Rennes, France, and cohead of the DiverSE research team. His research interests include model-driven and software language engineering and DevOps. Combemale received his a Ph.D. in software engineering from the University of Toulouse. He is a Member of IEEE. Contact him at benoit.combemale@irisa.fr and http://combemale.fr/.
https://orcid.org/0000-0002-7104-7848

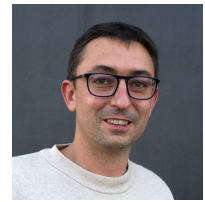

**Gunter Mussbacher** is an associate professor at McGill University, Montreal, QC H3A 0E9, Canada. His research interests include model-driven requirements and software language engineering, sustainability, and human values. Mussbacher received his Ph.D. in computer science from the University of Ottawa. Contact him at gunter.mussbacher@mcgill.ca and http://www.ece.mcgill.ca/~gmussb1/.
https://orcid.org/0009-0006-8070-9184

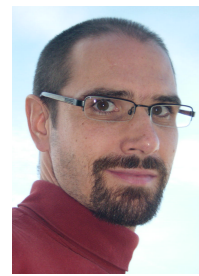

**PolyDebug: A Framework for Polyglot Debugging**

**Tijs van der Storm** is a senior researcher in the Software Analysis and Transformation (SWAT) group at Centrum Wiskunde & Informatica (CWI) in Amsterdam, and full professor in Software Engineering at the University of Groningen in Groningen. His research is centered around the questions how to make better programming languages and how to better make programming languages. Contact him at storm@cwi.nl. 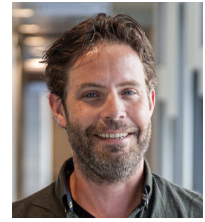
https://orcid.org/0000-0001-8853-7934